\documentstyle[12pt]{article}

\begin{document}

\LaTeX{}\bigskip\ \bigskip\ \bigskip\ \bigskip\ 

\begin{center}
A note on the thermal component of the equation of state of solids\medskip\ 

{\it (Letter to the Editor)}\bigskip\ \bigskip\ 

Vladan Celebonovic\medskip\ 

Institute of Physics,Pregrevica 118,11080 Zemun-Beograd,Yugoslavia\medskip\ 

(received: 15 November,1990)\bigskip\ 

reference: {\it Earth,Moon and Planets,}{\bf 54,} pp.145-149 (1991).\bigskip%
\ 
\end{center}

Abstract: A simple method for determining the thermal component of the EOS
of solids under high pressure is proposed.Applications to the interior of
the Earth gives results in agreement with recent geophysical data\bigskip\ 
\bigskip\ 

\begin{center}
{\bf 1.Introduction}
\end{center}

Imagine a specimen of any chemical substance.In the case of a
thermomechanical system,its equation of stae (EOS) will have the general
form $f(p,V,T)=0$,where $p,V,T$ denote,respectively the pressure to which
the system is subjected,its volume and temperature.

Proposing a realistic EOS for a given class of physical systems is,generally
speaking a highly non-trivial problem (see,for instance,Eliezer {\it et al.,}%
1986; Schatzman and Praderie,1990 for details ).One usually starts from an
assumed interparticle potential,determines the form of the Hamiltonian,and
from it the thermodynamical functions and the EOS.In order to render the
calculations more tractable,.the EOS is usually determined in form of
isoterms in the $p-V$ plane.The thermal component is often introduced at a
later stage (for example,Holian,1986;Eliezer {\it et al.,}1986).\newpage

In laboratory studies,EOS of various systems can be determined
experimentally (examples of recent reviews concerning the subject are
Jayaraman,1986;Jeanloz,1989;Drickamer,1990).The situation is much more
complicated in astrophysics,because planetary and stellar structure is
unaccessible to direct experiments.What is observable,however,are the
consequences on the surfaces and in the vicinity of these objects of the
processes occuring in their interiors.Progress in the understanding of the
Earth's interior has,for example,been provoked by a combination of
observation in geophysics and seismology,with laboratory studies of the
relevant materials performed under high pressure and high temperature
(Jeanloz,1990a).

The purpose of this letter is to propose a simple method for the
determination of the thermal component of the EOS of solids subdued to high
pressure.Details of the calculations,and and application to the interior of
the Earth are presented in the following section,while the third part
contains a discussion of various factors influencing our results.as well as
the possibilities of their improvement.Physically,our calculations are based
on combination known results of solid-state physics with a particular
semiclassical theory of dense matter proposed by Savic and Kasanin
(1962/65;for a recent application,see Celebonovic,1990b,and references given
there).Compared with recent work on the subject (such as Renero {\it et al.,}%
1990;Kumari and Dass,1990a,b),
the approach proposed in this letter has the advantage of physical and 
mathematical simplicity,but a disadvantage of being applicable only to 
solids. \bigskip\ 

\begin{center}
{\bf 2.Calculations}\bigskip\ 
\end{center}

One of the general characteristics of solid bodies is that the atoms (or
ions) in them,preform small vibrations about their equilibrium positions.Let 
$N$ be the number of molecules in the body,and denote by $\nu $ the number
of atoms per molecule.The total number of atoms in the system is $N\nu $ and
the number of vibrational degrees of freedom is $3N\nu $ (speaking
eaxctly,it is $3N\nu -6,$ but $3N\nu \gg 6$ ).\newpage 

A system of $3N\nu $ mutually independent vibrational degrees of freedom is
equivalent to an ensemble of independent oscillators.The free energy of such
an ensemble can be expressed as

\begin{equation}
\label{(1)}F=N\epsilon _0+k_BT\sum_\alpha \ln (1-\exp (-\beta \hbar \omega
_\alpha )) 
\end{equation}

where the first term on the right side represents the interaction energy of
all the atoms in the system in the equilibrium state,and the summation is
carried over all the normal vibrations,indexed by $\alpha $ .

At low temperatures,only low frequency terms (i.e.,sound waves) with $%
k_BT\cong \hbar \omega _\alpha $ have an important contirubution to
eq.(1).It can be shown (for example Landau and Lifchitz,1959) that the
number of vibrations $dN$ per interval of frequency $d\omega $ is

\begin{equation}
\label{(2)}\frac{dN}{d\omega }=\frac{3V\omega ^2}{2\pi ^2\overline{u}^3} 
\end{equation}

\medskip\ The volume of the system is denoted by V,and $\overline{u}$ is the
mean velocity of sound.Changing from summation to integration in eq.(1),one
finally obtains

\begin{equation}
\label{(3)}F=N\epsilon _0+\frac{3k_BTV}{2\pi ^2\overline{u}^3}\int_0^\infty
\ln (1-\exp (-\hbar \omega _\alpha /k_BT))d\omega 
\end{equation}

\medskip\ 

If we apply the standard thermodynamical identity $E=F-T\frac{\partial F}{%
\partial T}$ it follows from eq.(3) that the energy of the system is\medskip%
\ 

\begin{equation}
\label{(4)}E=N\epsilon _0+V\pi ^2(k_BT)^4/10(\hbar \overline{u}^3) 
\end{equation}

The mean velocity of sound can be approximated by the Bohm-Staver formula

\begin{equation}
\label{(5)}\overline{u}^2=(1/3)\left( Zm/Mv_F^2\right) 
\end{equation}

\medskip\ (Bohm and Staver,1951;Ashcroft and Mermin,1976),where $m$ and $M$
denote,respectively,the masses of electrons and ions in the solid,$v_F$ is
the Fermi velocity and $Z$ is the charge of ions. \newpage\ 

A detailed discussion of the basic ideas of the theory of Savic and Kasanin
(the SK theory for short)\ has recently been published (Celebonovic,1989d).
Within the framework of this theory,the energy per unit volume of a solid 
is given by

\begin{equation}
\label{(6)}E=2e^2Z(N_A\rho /A)^{4/3} 
\end{equation}

\medskip\ where $A$ is the mass number and $N_A$ denotes Avogadro's constant.

Using equations\ (4)-(6),after some algebra one obtains the following
expression for the temperature of a solid as a function of its density $\rho 
$

\begin{equation}
\label{(7)}T=1.4217*10^5(\rho /A)^{7/12}(m/M)^{3/8}Z^{7/8} 
\end{equation}

\medskip\ The thermal component of the EOS of a solid can be obtained by
multiplying this result by the density (expressed in suitable units).

As an astrophysical test,eq.(7) was applied to the model of the Earth
discussed previously within the SK theory (Savic,1981 and earlier work)
.The following results were obtained\bigskip\ 

\begin{center}
TABLE \quad I\medskip\ 
\end{center}

Depth (km) \qquad \negthinspace \negthinspace 0-39 \qquad  
\negthinspace 39-2900\qquad \negthinspace 2900-4980\qquad
4980-6371\smallskip\ 

$\rho _{\max }$(g cm$^{-3}$) \quad 3.0 \qquad 
\quad 6.0\qquad \qquad 12.0
\qquad\qquad 19.74 \smallskip\ 

$Z$ \qquad\qquad\qquad 2 \qquad\qquad 3
 \qquad\qquad\quad 3 \qquad\qquad\quad 4\smallskip\ 

$T_{\max }(K)$ \qquad\quad 1300 \qquad\quad 2700
\qquad\quad 4100 \qquad\quad 7000 \bigskip\ 

where the values of the temperatures have been rounded to the nearest $\pm
100K.$\newpage 

\begin{center}
{\bf 3.Discussion}\bigskip\ 
\end{center}

The distrribution of temperature with depth within the Earth,or any other
celestial body,is not directly measurable.In order to draw conclusions about
the thermodynamics of our planet's interior,geophysicists are bound to
combine examination of rock samples from the outher $\sim $ 200 km of the
Earth,with high pressure-high temperature experiments (Jeanloz,1990a,b).It
has thus been shown that,for example,$T=(4500\pm 500)K$ at the base of the
mantle,and that $T=(6900\pm 1000)K$ in the center (Jeanloz,1990a).

These experiments were performed on materials known to exist in a thin layer
beneath the Earth's surface,and assumed to represent its composition in
bulk.One could conjecture that,such an assumption not being directly
verifiable by experiments,it has strong influence on the results.Some
influence it certainly has,but it can not be extremely important.Namely,it
is possible within the SK theory,to determine the bulk chemical composition
of an astronomical object ({\it i.e.,} the mean value of the mass number of
the mixture of materials that it is made of);the only input data needed for
such a calculation are the mass and radius of the object.It turns out that
the value of A obtained in this way corresponds closely to the value of A
for the materials used in experiments.

The calculations discussed in this letter are strongly influenced by a
combination of several factors from the domain of solid state physics.

The upper limit of integration in eq.(3) is infinite,which simplifies
mathematics,but renders physics unrealistic - there are no infinite
frequencies.The difficulty could be,at first sight,circumvented by
introducing a suitable cut-ff frequency,in the spirit of Debye's
model.However,one would then encounter the following obstacle,which would be
the density dependence of the cut-off frequency (and the coresponding
cut-off temperature).Solving this problem demans a knowledge of the elastic
constans and inter-particle potentials of the material ( a solution for some
types of crystal lattices within Debye's model is given in de
Launay,1953,1954).\newpage\ 

It is clear that attempting to perform such a calculation for the case of
the interior of the Earth would quickly render the results questionable,due
to the accumulation of various approximations.Another ''solid-state'' factor
influencing our results is the method of calculation of the speed of
sound.The use of the Bohm-Staver formula amounts to taking into account only
the main term in the calculation of the band-structure energy of a solid
(see Harrison,1989 for details),which is ,in turn,used in determining the
value of the velocity of sound.The formula is known to give qualitatively
correct results when applied to materials under standard conditions,but its
applicability can be expected to increase for materials subjected to high
pressure.

Finally,a few words about one more ''influential'' factor.It could be
questioned whether it is correct to compare equations which do not contain
parameters of the same physical nature:equation (4) contains the
temperature,while eq.(6) does not.It is true that the SK theory has been
developed for the case T=0K.However,this should be understood just as a
simplifying assumption,whose physical meaning is that this theory is
applicable to materials at small teemperatures and subjected to high
pressure,for which $\epsilon _F/k_BT\succ 1$ (see Eliezer {\it et al.,}1986
for details).\bigskip\ 

\begin{center}
{\bf 4.Conclusions}
\end{center}

In this letter we have discussed a simple method for determining the thermal
component of the EOS of solids under high pressure,thus correcting a small
error made in a similar discussion (Celebonovic,1982).Application to the
interior of the Earth gives results in good agreement with geophysical
data.Possible influence of several factors on the values obtained has been
described in some detail.Using this method in laboratory high pressure work
would necessitate an improvement in the calculation of the velocity of
sound,and the intorduction of a suitable density-dependent cut-off frequency
in equation (3).\newpage\ 

\begin{center}
{\bf References}\medskip\ 
\end{center}

Ashcroft,N.W.,and Mermin,D.N.: 1976,Solid State Physics,Holt,Rinehart and
Winston,London.

Bohm,D.and Staver,T.:1951,Phys.Rev.,{\bf 84},836.

Celebonovic,V.:1982,in W.Fricke and G.Teleki (eds.),Sun and Planetary
System,D.Reidel Publ.Comp., Dordrecht,Holland.

Celebonovic,V.:1989d,Earth,Moon and Planets,{\bf 45},291.

Celebonovic,V.:1990b,High Pressure Research,{\bf 5},693.

de Launay,J.:1953,J.Chem.Phys.,{\bf 21},1975.

de Launay,J.:1954,J.Chem.Phys.,{\bf 22},1676.

Drickamer,H.G.:1990,Ann.Rev.Mater.Sci.,{\bf 20},1.

Eliezer,S.,Ghatak,A.and Hora,H.:1986,An Introduction to Equations of
State:Theory and Applications,Cambridge University Press,Cambridge,UK.

Harrison,W.A.:1989,Electronic Structure and the Properties

of Solids,Dover Publications Inc.,New York.

Holian,K.S.: 1986,J.Appl.Phys.,{\bf 59},149.

Jayaraman,A.:1986,Rev.Sci.Instr.,{\bf 57},1013.

Jeanloz,R.:1989,Ann.Rev.Phys.Chem.,{\bf 40},237.

Jeanloz,R.:1990a,Ann.Rev.Earth Planet.Sci.,{\bf 18},357.

Jeanloz,R.:1990b,preprint,to appear in Gibbs symposium Proceedings.

Kumari,M.and Dass,N.:1990a,J.Phys.:Condens.Matt.,{\bf 2},3219.

Kumari,M.and Dass,N.:1990b,ibid,7891.

Landau,L.and Lifchitz,E.:1959,Statistical Physics,

Pergamon Press,Oxford.

Renero,C.,Prieto,F.E.and de Icaza,M.:1990,J.Phys.:

Condens.Matt., {\bf 2},295.

Savic,P.and Kasanin,R.:1962/65,The Behaviour of Materials Under High
Pressure I-IV,Ed.SANU,Beograd.

Savic,P.:1981,Adv.Space Res.,{\bf 1},131.

Schatzman,E.and Praderie,F.:1990,Astrophysique:Les Etoiles,

InterEditions/Editions du CNRS,Paris.

\ \qquad

\end{document}